\documentclass[aps,prl,twocolumn,showpacs,superscriptaddress,groupaddress,floatfix,graphics,graphicx]{revtex4-1}
\usepackage[latin9]{inputenc}
\usepackage{color}
\usepackage{float}
\usepackage{amsmath}
\usepackage{amssymb}
\usepackage{graphicx}
\usepackage{wasysym}
\usepackage{subscript}
\usepackage[unicode=true,pdfusetitle,
 bookmarks=true,bookmarksnumbered=false,bookmarksopen=false,
 breaklinks=false,pdfborder={0 0 1},backref=false,colorlinks=false]
 {hyperref}
\hypersetup{
 colorlinks,linkcolor=blue,citecolor=blue,urlcolor=blue}

\makeatletter

\newcommand*\LyXThinSpace{\,\hspace{0pt}}
\DeclareFontEncoding{LGR}{}{}
\DeclareRobustCommand{\greektext}{%
  \fontencoding{LGR}\selectfont\def\encodingdefault{LGR}}
\DeclareRobustCommand{\textgreek}[1]{\leavevmode{\greektext #1}}
\ProvideTextCommand{\~}{LGR}[1]{\char126#1}

\makeatother

\begin{document}
\title{Unambiguous electrical detection of spin-charge conversion in lateral
spin-valves }
\author{Stuart A. Cavill}
\affiliation{Department of Physics, University of York, YO10 5DD, York, United
Kingdom}
\author{Chunli Huang}
\affiliation{Department of Physics, National Tsing Hua University and National
Center for Theoretical Sciences (NCTS), Hsinchu 30013, Taiwan}
\affiliation{Department of Physics, The University of Texas at Austin, Austin,
Texas 78712,USA}
\author{Manuel Offidani}
\affiliation{Department of Physics, University of York, YO10 5DD, York, United
Kingdom}
\author{Yu-Hsuan Lin }
\affiliation{Department of Physics, National Tsing Hua University and National
Center for Theoretical Sciences (NCTS), Hsinchu 30013, Taiwan}
\author{Miguel A. Cazalilla}
\affiliation{Department of Physics, National Tsing Hua University and National
Center for Theoretical Sciences (NCTS), Hsinchu 30013, Taiwan}
\author{Aires Ferreira}
\affiliation{Department of Physics, University of York, YO10 5DD, York, United
Kingdom}
\begin{abstract}
\textcolor{black}{Efficient detection of spin-charge conversion is
crucial for advancing our understanding of emergent phenomena in spin-orbit-coupled
nanostructures. Here, we provide proof of principle of an electrical
detection scheme of spin-charge conversion that enables full disentanglement
of competing spin-orbit coupling transport phenomena in diffusive
lateral channels i.e. the inverse spin Hall effect (ISHE) and the
spin galvanic effect (SGE). A suitable detection geometry in an applied
oblique magnetic field is shown to provide direct access to spin-charge
transport coefficients through a simple symmetry analysis of the output
non-local resistance. The scheme is robust against tilting of the
spin-injector magnetization, disorder and spurious non-spin related
contributions to the non-local signal, and can be used to probe spin-charge
conversion effects in both spin-valve and hybrid optospintronic devices. }
\end{abstract}
\maketitle
The generation and manipulation of nonequilibrium spins at interfaces
are key goals in the operation of spintronic devices, with recent
research heavily focused on using spin-orbit coupling (SOC) for achieving
both. \textcolor{black}{As the spin dynamics are} sensitive \textcolor{black}{to
symmetry-breaking effects, it is conspicuous that transport measurements
provide a powerful probe of emergent spin-orbit phenomena. As such},
a fundamental understanding of spin transport at interfaces is currently
a major enterprise towards the effective control over the spin degree
of freedom \citep{Review_Interfaces_15,Review_Interfaces_16,Review_Interfaces_17}. 

The conversion of spin current into a charge current is also essential
for an all-electrical readout. SOC has been shown or predicted to
provide a suitable means to achieve this in a number of systems with
broken inversion symmetry ranging from two-dimensional (2D) electron
gases in oxide-oxide \citep{1stP_SCC_2DEGOxide}, metal-semiconductor
\citep{1stP_SCC_Ge-metal} and metal-metal \citep{1stP_SCC_metal_metaloxide}
interfaces to surface states of topological insulators \citep{1stP_SCC_metal_TI}
and spin-split bulk states in polar semiconductors~\citep{1stP_BulkBReffect_BiTeI}.
More recently, bilayers of 2D crystals have emerged as highly-controllable
testbeds for exploring interfacial SOC phenomena due to their gate-tunable
charge carriers and interplay between spin and lattice-pseudospin
degrees of freedom \citep{GSOC_Rashba_09,GSOC_TMDs_Zhu_11,GSOC_TMDs_Xiao_12,GSOC_TMDs_Muniz_15,GSOC_Review_Garcia_18}.
This novel class of Dirac materials have further extended the breadth
of interfacial phenomena to encompass proximity-induced SOC within
atomically thin crystals \citep{GTMD_Avsar_14,GTMD_Wang_15,GTMD_Wang_16a,GTMD_Volkl_16b,GTMD_Yang_17,GTMD_Wakamura_18a,GTMD_Omar_18b},
all-optical spin-current injection \citep{GTMD_Optical_Luo_NanoLett2017,GTMD_Optical_Avsar_ACSNano2017},
large spin lifetime anisotropy \citep{GTMD_SRTA_Cummings,GTMD_SRTA_Ghiasi_17,GTMD_SRTA_Benitez18,GTMD_SRTA_Offidani_18}
and the co-existence of SHE and inverse SGE \citep{Milletari_ConservationLaws_17,GSOC_Offidani_17}.
Theoretical studies of 2D systems have also envisioned unconventional
spin-orbit scattering mechanisms, including robust skew scattering
from spin-transparent impurities \citep{Milletari_ConservationLaws_17}
and a direct magneto-electric coupling (DMC) effect---arising from
quantum interference between distinct components of the single-impurity
SOC potential---which generates a nonequilibrium spin polarization
\citep{GSOC_Huang_16}. 

The abundance of microscopic spin-orbit mechanisms in materials with
broken inversion symmetry motivates the search for device geometries
that can enable electrical detection of spin-charge interconversion
effects.\textcolor{black}{{} The H-bar scheme in Refs.\,\citep{GTMD_Avsar_14,GTMD_Wang_15}
employs the SHE for spin-current generation together with its Onsager
reciprocal, the inverse spin Hall effect (ISHE), for electrical read-out.
This method can be further extended to the detection of the SGE, the
Onsager-reciprocal of current-induced spin polarization, as shown
in Ref.\,\citep{Huang_NonLocalResistance_17}. However, the two-step
process, at the heart of the quadratic dependence of the nonlocal
resistance with the spin-charge conversion rate, has serious drawbacks.
Primarily it makes the H-bar approach prone to noise and variability,
as demonstrated by the experiments on graphene with adatoms \citep{AdGraph_Balakrishnan_13,AdGraph_Kaverzin_15,AdGraph_Volkl_19,AdGraph_Wang_15}.
Secondly, it precludes the unambiguous determination of the spin Hall
angle ($\theta_{\textrm{SHE}}$) and SGE efficiency ($\theta_{\textrm{SGE}}$)
in samples where these effects compete \citep{SHE_SGE_2D_1,SHE_SGE_2D_2,SHE_SGE_2D_3}.
Further progress in this flourishing field will require alternative
lateral-device detection schemes where competing SOC effects can be
simultaneously detected and quantified. }

\begin{figure}[t]
\includegraphics[width=1\columnwidth]{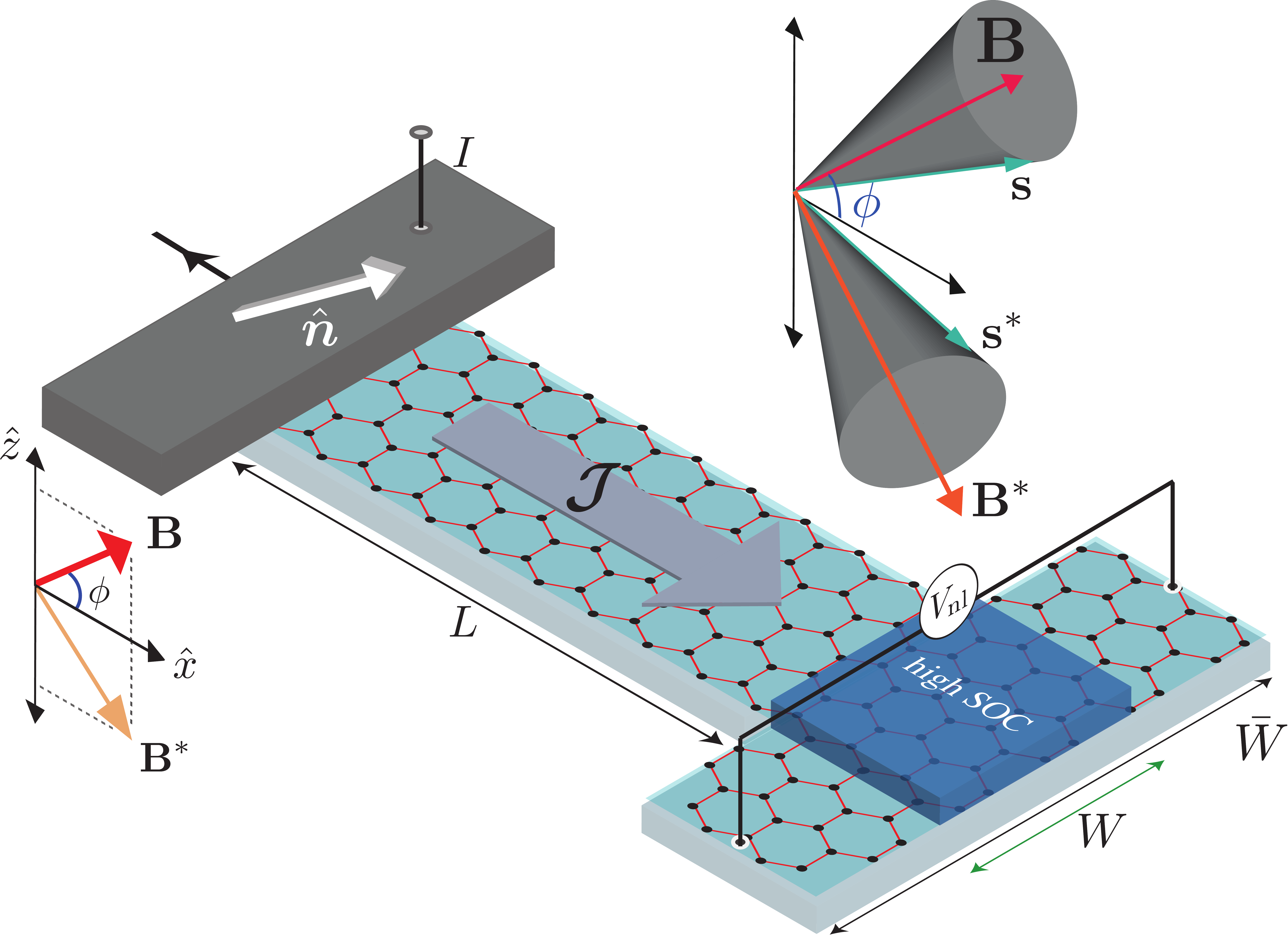} \caption{\label{fig:01}The injected non-equilibrium spin density $\mathbf{s}$
($\mathbf{s}^{*}$) parallel to the FM magnetization $\hat{n}=\mathbf{M}/|\mathbf{M}|$
diffuses through a cross-shaped channel under an applied oblique field
$\mathbf{B}$ ($\mathbf{B}^{*}$). Intrinsic and extrinsic SOC generate
a nonlocal voltage ($V_{\textrm{nl}}$).}
 
\end{figure}

In this Letter, we propose a measurement protocol for the \textit{unambiguous}
linear detection of ISHE and SGE in lateral spin\textcolor{black}{{}
devices. An opto-spintronic analogue, where the initial spin accumulation
is achieved by purely optical means, is also presented.} To illustrate
the general principle of the detection scheme, we focus on a nonlocal
setup comprising a spin-injector, a high-fidelity graphene spin-channel
\citep{SpinChannel_Tombros_07,SpinChannel_Kamalakar_15,SpinChannel_Gurram_18}
and a cross-shaped junction, where the graphene is covered by a high-SOC
material (see Fig.\,\ref{fig:01}). This layout is particularly well
suited to the measurement protocol because one can separate the spin-channel
from the SOC-active (spin-charge conversion) region. We stress, however,
that the detection scheme is \textit{general} to any lateral spin-injection
device providing that the breaking of inversion symmetry away from
the detection region is sufficiently weak, such that the channel length
($L)$ $\gtrsim$ spin diffusion length ($l_{s}$) $\gg$ mean free
path ($l$), which corresponds to the experimentally relevant diffusive
regime. As such, the proposed scheme can potentially shed light onto
spin-charge conversion effects in different material systems, such
as oxide heterostructures \citep{SID_Reyren}, metallic bilayers \citep{SID_Isasa}
and doped semiconductors \citep{SID_Kamerbeek,SID_Lou}, for which
spin injection has been recently established by nonlocal transport
methods, but the precise interplay of ISHE and SGE is yet to be uncovered. 

The general principle is akin to electrical detection of ISHE\,\citep{ISHE_Valenzuela}.
Spin-polarized carriers, injected by applying a current $I$ at a
ferromagnet (FM) contact, generate lateral charge accumulation via
SOC, which is then detected as a nonlocal voltage at the Hall cross.
The possibility unveiled here to isolate SGE and ISHE contributions
on the output signal hinges on a fundamental distinction between the
non-equilibrium spin-polarization density induced Hall effect (SGE)
and the most familiar spin-current-induced Hall effect (ISHE). In
the former, the component of the diffusive  spin accumulation, with
spin moment \emph{collinear} to the propagation direction, generates
a transverse charge current $J_{i}=\epsilon_{ij}\,\theta_{\textrm{SGE}}\,\mathcal{J}_{j}^{j}\:$($i,j=x,y$),
where $\mathcal{J}_{i}^{a}$ is the spin current density in direction
$i$ with the spin polarized in $a$ ($a=x,y,z$). Conversely, in
the ISHE, the spin-moment, the spin current density and the charge
current density are mutually \emph{orthogonal} such that $J_{i}=\epsilon_{ijk}\,\theta_{\textrm{SHE}}\,\mathcal{J}_{j}^{k}$,
where $\epsilon_{ij}$ ($\epsilon_{ijk}$) stands for the Levi-Civita
symbol and summation over repeated indices is implied. Note that the
steady-state spin current is linked to the spin polarization in the
channel as $\mathcal{J}_{i}^{a}=-D_{s}\partial_{i}s^{a}$, where $D_{s}$
is the diffusion constant (a more accurate definition is given below).
The crucial role played by the active polarization channel---out-of-plane
($s^{z}$) for ISHE and in-plane ($s^{x}$) for SGE---is borne out
when coherent spin precession is induced by an \emph{oblique} magnetic
field normal to the spin-injector easy-axis, $\mathbf{B}=(B_{x},0,B_{z})$,
forcing diffusive spins to undergo ISHE and SGE \emph{simultaneously}
(Fig.~\ref{fig:01}). Oblique spin precession is a powerful probe
of spin relaxation anisotropy \citep{Raes_SRTA_Graphene_16,GTMD_SRTA_Ringer}
and here we show that a suitable detection scheme in applied oblique
field provides access to the charge-spin transport coefficients $\theta_{\textrm{SHE}}$
and $\theta_{\textrm{SGE}}$. Our main result is a linear filtering
protocol for the ISHE ($+$)/SGE($-$) nonlocal resistance 
\begin{equation}
R_{\textrm{ISHE(SGE)}}=\frac{1}{2}\,[\Delta R_{\textrm{nl}}(\mathbf{B})\pm\Delta R_{\textrm{nl}}(\mathbf{B}^{*})]\,,\label{eq:main_finding}
\end{equation}
where $\Delta R_{\textrm{nl}}=\frac{1}{2I}(V_{\textrm{nl};n_{y}>0}-V_{\textrm{nl};n_{y}<0})$,
with $\hat{\boldsymbol{n}}=\mathbf{M}/|\mathbf{M}|$, is the output
transresistance difference between opposite configurations of the
spin-injector and $\mathbf{B}^{*}=(B_{x},0,-B_{z})$ is the mirror-image
of $\mathbf{B}$. The spin-charge conversion efficiency parameters
($\theta_{\textrm{SHE}}$, $\theta_{\textrm{SGE}}$) can be accurately
determined using a simple model of the spin-injector magnetization
tilting $\mathbf{M}(\mathbf{B})$. (For optical spin-injection, the
detection is carried out with standard Hanle technique.)

\emph{Theory}.---We give an intuitive proof of Eq.\,(\ref{eq:main_finding}).
In the narrow channel limit ($W\ll l_{s}$), the spin dynamics of
a typical disordered sample with weak SOC ($l_{s}\gg l$) are well
captured by the generalized 1D Bloch model $\partial_{t}\mathbf{\bar{s}}=D_{s}\partial_{x}^{2}\mathbf{\bar{s}}+\gamma\,(\mathbf{\bar{s}}\times\mathbf{B})-\hat{\Gamma}\,\mathbf{\bar{s}}$
\citep{Raes_anisotropic}, where $\mathbf{\bar{s}}\equiv(\mathbf{s}_{n_{y}>0}-\mathbf{s}_{n_{y}<0})/2$
is the spin-density difference between opposites configurations of
the spin-injector, $\hat{\Gamma}=\textrm{diag}\{1/\tau_{s}^{x},1/\tau_{s}^{y},1/\tau_{s}^{z}\}$
and $\gamma$ is the gyromagnetic ratio. The spin-relaxation matrix
$\hat{\Gamma}$ encapsulates the effects of spin-dephasing \citep{Cummings_16}
and irreversible spin relaxation mechanisms (which need not to be
isotropic due to SOC \citep{GTMD_SRTA_Offidani_18}). The output signal
$\Delta R_{\textrm{nl}}$ is proportional to the transverse current
generated at the cross. A simple dimensional analysis yields: $\Delta R_{\textrm{nl}}\propto D_{s}\,[\theta_{\textrm{SHE}}\partial_{x}\bar{s}^{z}(x)+(\theta_{\textrm{SGE}}/l_{s})\bar{s}^{x}]_{x=L}$.
Crucially, since the injected spins follow the contact magnetization,
the boundary term satisfies $\mathbf{\bar{s}}(x=0)||\hat{y}$, for
any $\mathbf{B}\perp\hat{y}$. This implies that the solution of the
1D Bloch equation for the spin polarization density \emph{difference}
$\bar{\mathbf{s}}$ transforms as $(\bar{s}^{x}(x),\bar{s}^{z}(x))\rightarrow(-\bar{s}^{x}(x),\bar{s}^{z}(x))$,
under the operation $B_{z}\rightarrow-B_{z}$. This shows that $\Delta R_{\textrm{nl}}(\mathbf{B})=\Delta R_{\textrm{nl}}(\mathbf{B}^{*})$
for SHE and $\Delta R_{\textrm{nl}}(\mathbf{B})=-\Delta R_{\textrm{nl}}(\mathbf{B}^{*})$
for SGE, thus proving the generality of the filtering scheme Eq.\,(\ref{eq:main_finding}). 

We formalize our results with a theory of coupled spin-charge diffusive
transport for disordered 2D conductors subject to random SOC sources
\citep{Lin_LongPaper}. The charge and spin observables, coarse-grained
over typical distances longer than the mean free path, satisfy the
continuity-like equation: 
\begin{equation}
[\nabla_{t}s]^{a}+\big[\nabla_{i}\mathcal{J}_{i}\big]^{a}=-\frac{s^{a}}{\tau_{s}^{a}}-\kappa_{i}^{a}J_{i},\label{eq:s-cont}
\end{equation}
the continuity relation $\partial_{t}\rho=-\partial_{i}J_{i},$ and
the generalized constitutive relations
\begin{equation}
J_{i}=-D_{c}\,\left(\partial_{i}\rho+\kappa_{i}^{a}s^{a}\right)+\gamma_{ij}^{a}\mathcal{J}_{j}^{a},\label{eq:c-consti}
\end{equation}
\begin{equation}
\mathcal{J}_{i}^{a}=-D_{s}\left[\nabla_{i}s\right]^{a}+\gamma_{ij}^{a}J_{j}.\label{eq:s-consti}
\end{equation}
The coefficients appearing in Eqs.\,(\ref{eq:s-cont})-(\ref{eq:s-consti})
can be obtained from microscopic calculations \citep{Lin_LongPaper,Shen_PRB2014,Burkov_04,Tokatly_Sherman_10}.
The DMC efficiency $\kappa_{i}^{a}$ (units $m^{-1}$) and spin Hall
angle $\gamma_{ij}^{a}$ quantify the spin-charge conversion efficiency
of random SOC sources. In this compact formalism, the effects of all
\textcolor{black}{spatially uniform spin-dependent interactions }are
captured by the SU(2) gauge field\textcolor{black}{{} $\mathcal{A}_{\mu}=A_{\mu}^{a}\,\sigma^{a}$
($\mu=t,i$),} with $\sigma^{a}$\textcolor{black}{{} }the spin-1/2
Pauli matrice\textcolor{black}{s. Th}e gauge field is implicit in
the covariant derivative \textcolor{black}{$[\nabla_{\mu}O]^{a}=\partial_{\mu}O^{a}\pm\epsilon^{abc}A_{\mu}^{b}O^{c}$,
with $O^{a}=\{s^{a},\mathcal{J}^{a}\}$ a spin observable and the
sign $\pm$ holding for a space ($+$)/time($-$) derivative. The
time component $A_{0}^{a}=g\mu_{B}B^{a}$ reproduces the spin precession
in the external Zeeman field and $A_{i}^{a}$ encodes the symmetry-breaking
SOC \citep{Tokatly_08}. We specialize to uniform SOC of the Bychkov-Rashba
(BR) form, which is ubiquitous in interfaces with broken inversion
symmetry. For 2D Dirac fermions, its non-zero components read as $A_{y}^{x}=-A_{x}^{y}=\lambda/v_{F}$,
with $v_{F}$ the Fermi velocity and $\lambda$ the BR-spin gap, whereas
for parabolic 2D electron gases, $A_{y}^{x}=-A_{x}^{y}=2m^{*}\alpha$,
with $m^{*}$ the effective mass and $\alpha$ the BR coupling. }The
BR effect contributes to the total spin-charge conversion rates with
terms proportional to the momentum scattering time $\tau$, namely
(i) an ISHE-like term $\gamma_{ij}^{a;\textrm{BR}}=\epsilon_{ij}\,\omega_{\textrm{BR}}\tau$,
where $\omega_{\textrm{BR}}=\lambda^{2}/\epsilon$ with $\epsilon$
the Fermi energy (2D Dirac gas) and $\omega_{\textrm{BR}}=2m\alpha^{2}$
(2D electron gas); and (ii) an indirect coupling between spin polarization
density and charge current that modifies the DMC coefficient according
to $\kappa_{i}^{a}\rightarrow\kappa_{i}^{a}-\epsilon_{ia}\omega_{\textrm{BR}}\tau\,l_{R}^{-1}$,
where $l_{R}\equiv\hbar/|A_{x}^{y}|$ is the length scale of spin
precession about the BR field. 

Armed with this formalism, we can evaluate the spin density profile
in applied oblique field and establish the range of validity of the
linear filtering protocol {[}Eq.~(\ref{eq:main_finding}){]}. Beyond
the SU(2) gauge field and impurity scattering, we note that the spin-charge
conversion rates can also receive contributions from Berry curvature
and phonons \citep{RMP_SHE,Gorini_15}.  For ease of notation, we
lump all these contributions under the effective ISHE and SGE rates
$\gamma_{ij}^{a}=\theta_{\textrm{SHE}}\epsilon_{ij}\delta_{az}$ and
$\kappa_{i}^{a}=l_{\textrm{SGE}}^{-1}\epsilon_{ia}$, where $\delta_{ij}$
is the Kronecker delta symbol\emph{. }For disordered samples in the
typical weak SOC regime $(l_{s}/l_{\textrm{SGE}},\ \theta_{\textrm{SHE}})\ll1$,
the build up of a nonlocal voltage in the cross-shaped device involves
two independent steps as described below. 

\emph{Spin} \emph{precession} \emph{and spin-charge conversion}.---The
effective Bloch equation governing the steady-state spin accumulation
is obtained from Eqs.\,(\ref{eq:s-cont})-(\ref{eq:s-consti}) after
eliminating the spin current in favor of the spin polarization density.
To leading (linear) order in the spin-charge conversion rates, we
find
\begin{equation}
\boldsymbol{\mathcal{\bar{D}}}\cdot\mathbf{s}(x)+\gamma\left(\mathbf{s}\left(x\right)\times\boldsymbol{B}\right)=0\,,\label{eq: Diffusion equation}
\end{equation}
with \textcolor{black}{$\gamma=g\mu_{B}/\hbar$ and }
\begin{equation}
\boldsymbol{\mathcal{\bar{D}}}=D_{s}\left(\begin{array}{ccc}
\partial_{x}^{2}-l_{s,x}^{-2} & 0 & \kappa_{R}\partial_{x}\\
0 & \partial_{x}^{2}-l_{s,y}^{-2} & 0\\
-\kappa_{R}\partial_{x} & 0 & \partial_{x}^{2}-l_{s,z}^{-2}
\end{array}\right)\,\label{eq:diff_operator}
\end{equation}
where $l_{s,a}=\sqrt{D_{s}\tau_{s}^{a}}$ ($a=x,y,z$). The diffusion
matrix $\boldsymbol{\mathcal{\bar{D}}}$ displays the standard Fick
terms (diagonal), accounting for spin relaxation processes \citep{Raes_anisotropic},
and a conspicuous nondiffusive term proportional to the wavelength
of inversion symmetry breaking within the 1D channel $\kappa_{R}=2l_{R}^{-1}$.\textcolor{black}{{}
This term is usually neglected in the analysis of spin precession
\citep{Raes_anisotropic}, since it is assumed that the main effect
of SOC is a renormalization of spin lifetimes. However, the more general
Bloch equation (\ref{eq: Diffusion equation}) shows that the BR gauge
field effectively induces a coherent coupled precession of $s_{x}$
(SGE) and $s^{z}$ (ISHE) components in materials with $\kappa_{R}\gtrsim l_{s}^{-1}$.
The protocol, which relies on the decoupling of SGE and ISHE precessions
using suitable oblique fields $\mathbf{B},\mathbf{B}^{*}\perp\hat{y}$,
is essentially exact for $l_{R}\gg l_{s}$, since in this limit Eq.
(\ref{eq: Diffusion equation}) reduces to its standard diffusive
form. Remarkably, the scheme is still accurate in channels with moderate
to strong BR effect, where $l_{R}$ can be only few times larger than
$l_{s}$; a detailed discussion is given in the supplemental material
(SM) \citep{SM}.} 

We now turn our attention to the detection region, where SOC generates
a transverse current $J_{y}$. The nonlocal voltage build-up is determined
by the open circuit condition $J_{y,\textrm{total}}=J_{y}+(\sigma_{\textrm{2D}}/\bar{W})V_{\textrm{nl}}=0$,
where $\sigma_{\textrm{2D}}$ is the DC conductivity. From Eqs.\,(\ref{eq:c-consti})-(\ref{eq:s-consti}),
we find
\begin{equation}
V_{\textrm{nl}}(x)=-\frac{\bar{W}D_{s}}{\sigma_{\textrm{2D}}}\left[l_{\textrm{SGE}}^{-1}\,s^{x}\left(x\right)+\theta_{\textrm{SHE}}\,\partial_{x}s^{z}\left(x\right)\right]+\mathcal{O}\,,\label{eq:Rnl}
\end{equation}
where $\mathcal{O}$ are next order corrections in $\theta_{\textrm{SHE}}$
and $l_{\textrm{SGE}}^{-1}$, thus validating the expression of $\Delta R_{nl}(x)$
obtained earlier via a dimensional analysis argument. 

\begin{figure}
\centering{}\includegraphics[width=0.9\columnwidth]{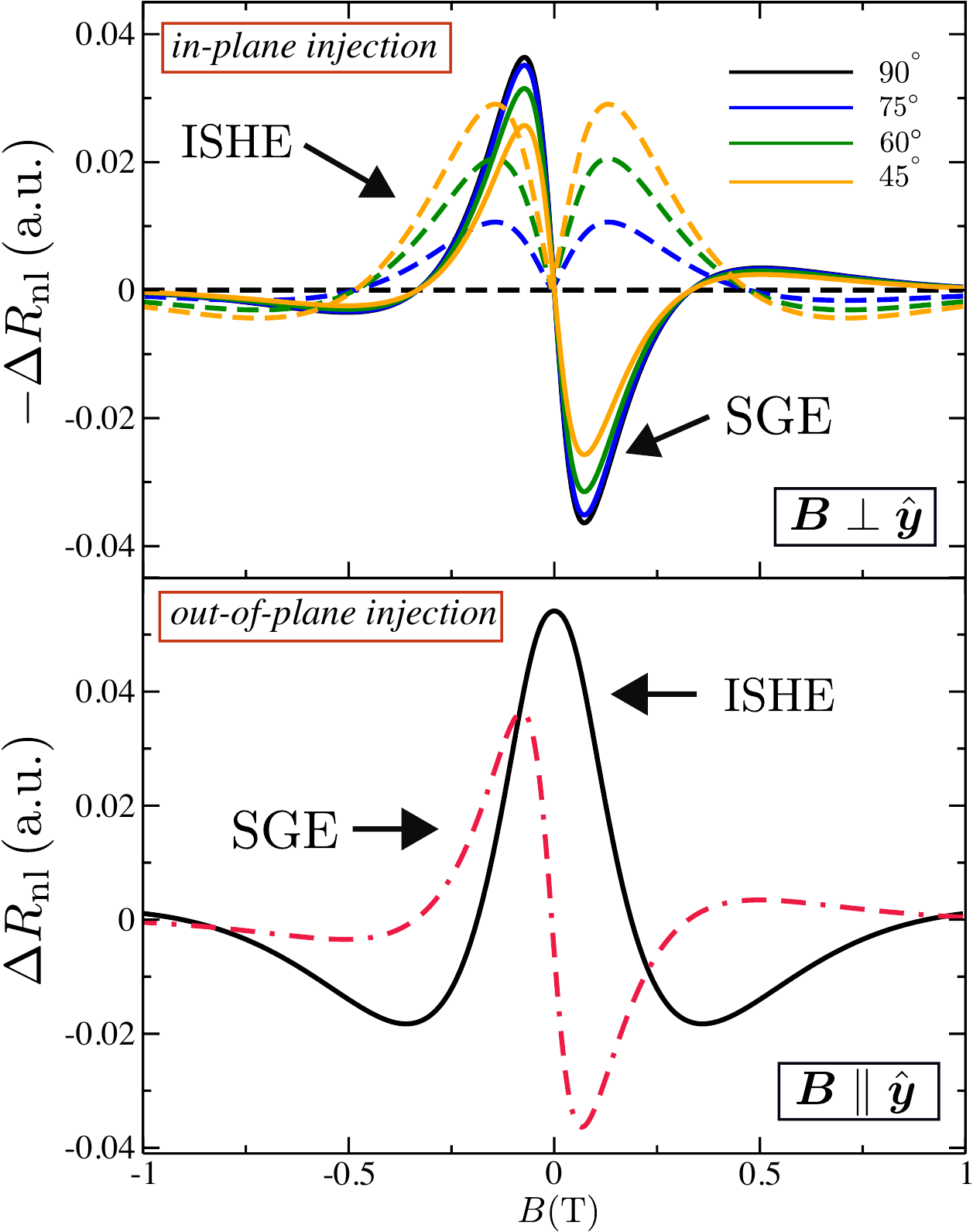}\caption{\label{fig:02}Detection of ISHE/DMC contributions for in-plane and
out-of-plane spin injection. Oblique field parameterized as $\boldsymbol{B}=|B|\cos\phi\thinspace\hat{\boldsymbol{x}}+B\sin\phi\thinspace\hat{\boldsymbol{z}}$
(see text for alternative measurement protocol with $\boldsymbol{B}\perp\hat{\boldsymbol{x}}$).
Results are for a typical graphene channel ($L=2l_{s}$ and $\tau_{s}=0.1\textrm{\text{\,}ns}$)
and heterojunction with $\theta_{\textrm{sH}}=0.01$ and $l_{\textrm{DMC}}=100l_{s}$. }
\end{figure}

\emph{Detection}.---We will now show how the detection is carried
out in a real setup. We first consider the spin-valve layout in Fig.
\ref{fig:01}. The injected spin polarization is parallel to the spin-injector
quantization axis (defined by $\mathbf{M}$) and thus is very sensitive
to the applied field \citep{SpinChannel_Gurram_18}. First, a field
$\mathbf{B}_{\textrm{in}}\parallel\pm\hat{y}$ is applied to set the
injector configuration either ``up'' ($n_{y}>0$) or ``down''
($n_{y}<0$). Subsequently, the field is removed and an oblique field
in the $Oxz$ plane is swept, $\boldsymbol{B}=(|B|\cos\phi\thinspace\hat{\boldsymbol{x}}+B\sin\phi\thinspace\hat{\boldsymbol{z}})$,
across first and fourth quadrants ($\phi\in[0,90^{\circ}]$). The
nonlocal resistance difference between opposite configurations of
the spin injector, $\Delta R_{\textrm{nl}}\equiv[\left.V_{\textrm{nl}}(L)\right|_{n_{y}>0}-\left.V_{\textrm{nl}}(L)\right|_{n_{y}<0}]_{x=L}/2I$
is obtained from Eq.\,(\ref{eq:Rnl}) upon solving the effective
spin Bloch equation {[}Eq.\,(\ref{eq: Diffusion equation}){]} for
the spin density profile $s^{a}(x)$. After a straightforward calculation,
we find 
\begin{equation}
\Delta R_{\textrm{nl}}=R_{0}\,\textrm{Im}\,\left[e^{-qL}\left(\theta_{\textrm{SHE}}ql_{s}s_{B}\cos\phi-\frac{l_{s}}{l_{\textrm{SGE}}}\sin\phi\right)\right]\,,\label{eq:deltaRn}
\end{equation}
where $q=l_{s}^{-1}\sqrt{1+\imath B\gamma\tau_{s}}$ is the complex
spin-precession wavevector, $s_{B}=\textrm{sign}(B)$, and $R_{0}=s^{y}(0)(\bar{W}D_{s}/l_{s}I\sigma_{\textrm{2D}})$.
We have also assumed an isotropic spin lifetime $\tau_{s}^{a}=\tau$,
which is characteristic of single-layer graphene \citep{Raes_SRTA_Graphene_16}.
The two contributions in Eq.\,(\ref{eq:deltaRn}) have opposite parity
under mirror reflection $B\rightarrow-B$ (or equivalently, $\phi\rightarrow-\phi$).
This in turn implies that the \emph{bona fide} ISHE ($+$)/SGE($-$)
spin-transresistance satisfies

\begin{equation}
\left.\Delta R_{\textrm{nl}}\right|_{\textrm{ISHE(SGE)}}^{\boldsymbol{B}\perp\hat{\mathbf{y}}}=\frac{1}{2}\left(\left.\Delta R_{\textrm{nl}}\right|_{B}\pm\left.\Delta R_{\textrm{nl}}\right|_{-B}\right),\label{eq:sym_asym_contrib}
\end{equation}
which is simply a special case of Eq. (\ref{eq:main_finding}). The
signal subtraction $\Delta R_{\textrm{nl}}\equiv\frac{1}{2}(\left.R_{\textrm{nl}}\right|_{n_{y}>0}-\left.R_{\textrm{nl}}\right|_{n_{y}<0})$
ensures that non-spin-related contributions, such as the local Hall
effect due to stray fields generated by the FM contact, are filtered
out. Moreover,\textcolor{black}{{} the spin-transresistance scale $R_{0}\propto s^{y}(0)\equiv s^{y}(0,\mathbf{B})$
defines the maximum achievable nonlocal-signal. Contacts with low
magnetic anisotropy saturate easily ($R_{0}\propto s^{y}(0,\mathbf{B})\rightarrow0$
for $B_{x,z}\gg B_{x,z}^{\textrm{sat}}$), which shrinks the $\Delta R_{\textrm{nl}}$-lineshape
with respect to an ideal spin-injector with magnetization pinned along
the easy axis $\mathbf{M}\parallel\pm\hat{y}$.} The typical lineshapes
for an isotropic channel with an ideal contact ($\hat{\mathbf{n}}||\pm\hat{\mathbf{y}}$)
are shown in Fig.\,\,\ref{fig:02}(a). For field applied perpendicularly
to the plane ($\phi=90^{\circ}$), spins precess in the plane, there
is no available out-of-plane spin density and thus ISHE is absent,
while SGE achieves is highest magnitude. When the field is tilted
towards the 2D plane ($\phi<90^{\circ}$), a symmetric ISHE component
appears. The realistic lineshapes are obtained by a trivial rescaling
$\Delta R_{\textrm{nl}}\rightarrow\mathcal{V}_{B}\,\Delta R_{\textrm{nl}}$,
where $\mathcal{V}_{B}=s^{y}(0,B)/s^{y}(0,0)$ is the ``visibility''
factor that takes into account response of the\textcolor{black}{{} spin-injector}
\citep{SM}. As aforementioned, since $\mathcal{V}_{B}=\mathcal{V}_{-B}$
(for a homogeneous contact) the only effect of the FM response within
the protocol is a shrinkage of the filtered signals. We briefly discuss
an alternative  detection scheme with an oblique field applied in
the $Oyz$-plane \citep{Raes_SRTA_Graphene_16}. The nonlocal resistance
for a fixed spin-injector configuration, $R_{\textrm{nl}}^{\textrm{YZ}}=V_{\textrm{nl}}^{\textrm{YZ}}/I$,
is found as
\begin{align}
R_{\textrm{nl}}^{\textrm{YZ}} & =r_{0}\,\left[\theta_{\textrm{SHE}}\sin(2\phi)f(q)-\frac{l_{s}\sin\phi}{l_{\textrm{SGE}}}\textrm{Im}\thinspace e^{-qL}\right]\,,\label{eq:Rnl_YZ}
\end{align}
where $r_{0}=[s^{y}(0)]_{\mathbf{B}}\times(\bar{W}D/l_{s}I\sigma_{\textrm{2D}})$
is the typical transresistance factor and $f(q)=[e^{-L/l_{s}}-Re\left(ql_{s}e^{-qL}\right)]/2$.
The ISHE/SGE components can be extracted via Fourier analysis if the
$\phi$-dependence of $r_{0}$ is known with good resolution. Alternatively,
for typical channels ($L\gtrsim l_{s}$), $\theta_{\textrm{SHE}}$
can be determined directly from nonlocal signal saturation at large
field $R_{\textrm{nl}}^{YZ}\rightarrow(r_{0}/2)\theta_{\textrm{SHE}}\sin(2\phi)e^{-L/l_{s}}$.
Once $\theta_{\textrm{SHE}}$ is determined, the SGE coefficient is
easily retrievable by fitting the full lineshape to Eq.\,(\ref{eq:Rnl_YZ}). 

We now discuss an optospintronic analogue, in which a transition metal
dichalcogenide monolayer replaces the FM contact as a spin injector
\citep{GSOC_TMDs_Muniz_15}. Optical spin-injection has been recently
demonstrated in graphene heterostructures \citep{GTMD_Optical_Avsar_ACSNano2017,GTMD_Optical_Luo_NanoLett2017}.
Since optically-pumped spin currents from atomically thin semiconductors
with spin-valley coupling are polarized normal to the 2D plane, an
in-plane field $\boldsymbol{B}=B\hat{y}$ can be used to detect ISHE
and SGE simultaneously. The spin-transresistance $\Delta R_{\textrm{nl}}^{z}\equiv[\left.V_{\textrm{nl}}(L)\right|_{n_{z}>0}-\left.V_{\textrm{nl}}(L)\right|_{n_{z}<0}]_{x=L}/2I$
is easily computed as
\begin{align}
\Delta R_{\textrm{nl}}^{z} & =r_{0}^{z}\,\textrm{Re}\left[\left(ql_{s}\theta_{\textrm{SHE}}-\imath l_{\textrm{SGE}}^{-1}\right)e^{-qL}\right]\,,\label{eq:NL_resistance_special_case}
\end{align}
where $q$ is defined as before and $r_{0}^{z}=s^{z}(0)\bar{W}D/(l_{s}I\sigma_{\textrm{2D}})$.
Similar to in-plane spin injection configuration, the ISHE and SGE
signals have different parity under mirror reflection $B\rightarrow-B$.
Typical lineshapes are shown in Fig.\,\ref{fig:02}(b). 

\emph{Experimental feasibility.} The measurement protocol is directly
applicable to devices with anisotropic spin dynamics, e.g. due to
BR effect or spin-valley locking \citep{Yang15}. Different material
systems and device layouts can be used providing (i) the injection/detection
regions are well separated to avoid the use of large precession fields
($L\gtrsim l_{s}$) and (ii) the inversion symmetry breaking within
the spin channel is weak enough ($l_{R}\apprge l_{s}$). \textcolor{black}{Such
a moderate SOC regime also guarantees that the spin-charge conversion
is within the linear regime (Eq.~(\ref{eq:Rnl})), which is indeed
observed for most devices }\citep{SID_Reyren,SID_Isasa,SID_Lou,SID_Kamerbeek,SpinChannel_Gurram_18,SpinChannel_Kamalakar_15,SpinChannel_Tombros_07,SHE_SGE_2D_1,SHE_SGE_2D_2,SHE_SGE_2D_3}\textcolor{black}{.
For example, in monolayer graphene, one finds $l_{R}\approx\hbar v_{F}/\lambda\approx14$
$\mu\textrm{m}\gg l_{s}$ for a typical $\lambda=50$ $\mu$eV} \citep{Min_06,Konschuh_10}.
For interfaces hosting 2D electron gases with weak or moderate BR
effect, the situation is similar. For example, using representative
values for the Au(111) surface, $\alpha=$$0.03$ eV$\cdot\mathring{\textrm{A}}$
and $m^{*}=0.26m_{e}$ \citep{Popovic05}), one obtains $l_{R}\approx100$
nm $\gg l_{s}$. \textcolor{black}{Beyond this regime, the interpretation
of experimental data requires explicit modeling (using a self-consistent
solution of Eqs. (\ref{eq:s-cont})-(\ref{eq:s-consti})), }hindering
the unambiguous detection of ISHE/SGE. 

We conclude with a remark on the spin-injector magnetization response
to the applied field. As shown in this work, the tilting of the magnetization
$\mathbf{M}$ away from its preferred easy axis reduces the ISHE/SGE
nonlocal resistance by a nonlinear factor (``visibility'') given
by $\mathcal{V}_{\mathbf{B}}=[s^{y}(0)]_{\mathbf{B}}/[s^{y}(0)]_{\textrm{ideal}}$.
The field dependence of the initial spin accumulation $[s^{y}(0)]_{\mathbf{B}}$
can be accurately described by a Stoner-Wohlfarth model with parameters
determined from separate Hanle  measurements. According to the realistic
simulations provided in the SM \citep{SM}, the visibility can be
as large as 20\% at low fields $\sim0.1$ T for typical FM saturation
parameters. We expect that the measurement scheme introduced in this
work will enable accurate experimental determination of spin-charge
conversion parameters ($\theta_{\textrm{SHE}}$, $\theta_{\textrm{SGE}}=l_{s}/l_{\textrm{SGE}}$).
We note that a similar scheme can in principle be employed for electrical
detection of spin Hall effect and Edelstein effect, the Onsager reciprocal
phenomena of ISHE and SGE. For example, this could be achieved by
injecting a current perpendicular to the channel and measuring the
resulting spin-electrochemical accumulation at the FM contact.\textcolor{blue}{{} }

In compliance with EPSRC policy framework on research data, this publication
is theoretical work that does not require supporting research data.

\emph{Acknowledgements.--}A.F. gratefully acknowledges the financial
support from the Royal Society, London through a Royal Society University
Research Fellowship. M.O. and A.F. acknowledge funding from EPSRC
(Grant Ref: EP/N004817/1). Y.-H. L., C. H. , and M.A.C. acknowledge
support from the Ministry of Science and Technology of Taiwan through
grants 102-2112-M-007- 024-MY5 and 107-2112-M-007-021-MY5, as well
as from the National Center for Theoretical Sciences (NCTS, Taiwan).

\newpage{}

\newpage{}

\onecolumngrid

\section*{S1. Nonlocal Resistance: Injector Magnetization Tilting }

In this section, we illustrate the application of the filtering protocol
with realistic simulations of the nolocal resistance that take into
account the FM magnetization tilting. The $\mathbf{B}$-dependent
polarization $\hat{\boldsymbol{n}}$ of injected spins is parameterized
as follows
\begin{equation}
\hat{\boldsymbol{n}}(B,\phi)=\hat{x}\cos\left[\gamma(B,\phi)\right]\sin\left[\beta(B,\phi)\right]+p\,\hat{y}\cos\left[\gamma(B,\phi)\right]\sin\left[\beta(B,\phi)\right]+\hat{z}\sin[\gamma(B,\phi)]\,,\label{eq:aux_1}
\end{equation}
where $p=\pm1$ for initial $\hat{\boldsymbol{n}}=$$\pm\hat{y}$
magnetic configurations of the ferromagnetic injector. The field-dependence
of the angle variables ($\beta,\gamma$) is usually described by a
Stoner-Wohlfarth model \citep{Raes_SRTA_Graphene_16,SpinChannel_Tombros_07}.
For our purposes, it suffices to consider a heuristic model for the
tilting angles
\begin{equation}
\beta(B,\phi)=\arctan(B_{x}/B_{\textrm{sat;x}})\,,\qquad\gamma(B,\phi)=\arctan(B_{z}/B_{\textrm{sat;z}})\,,\label{eq:aux_2}
\end{equation}
where $B_{\textrm{sat;}i}$ is the saturation field normal to the
easy axis $\hat{i}\perp\hat{y}$ ($B_{\textrm{sat;z}}\gg B_{\textrm{sat;x}}$
due to shape anisotropy~\citep{SpinChannel_Tombros_07,Raes_SRTA_Graphene_16}).
\smallskip{}

\begin{figure}[b]
\centering{}\includegraphics[width=1\textwidth]{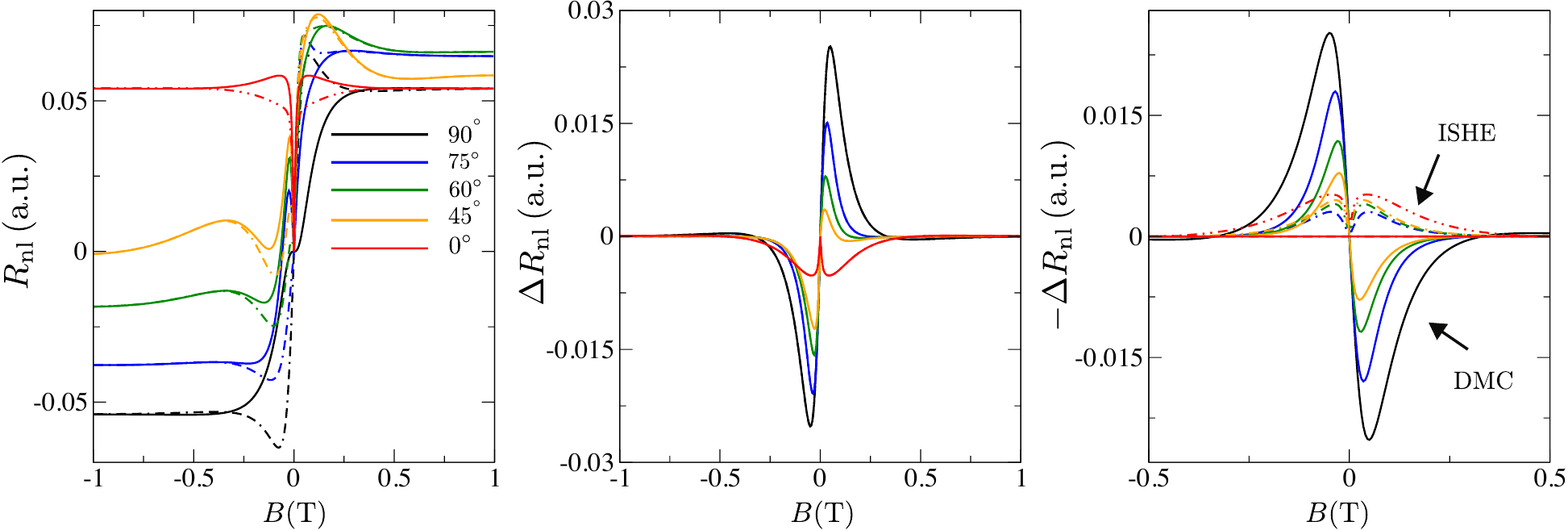} \caption{\label{fig:supp_1}Extraction of ISHE/SGE contributions to the spin
transresistance at selected oblique fields ($\phi=45^{\circ},60^{\circ}$),
in-plane field ($\phi=0^{\circ}$) and perpendicular field ($\phi=90^{\circ}$)
for spin-valve device comprising an isotropic graphene channel and
a high-SOC heterojunction. \textbf{Left}: Bare signal $R_{\textrm{nl}}$
for initial parallel (solid) and antiparallel (dashed) magnetic configurations.
\textbf{Middle}: Nominal nonlocal resistance $\Delta R_{\textrm{nl}}=[\Delta R_{\textrm{nl}}(n_{y}>0)-\Delta R_{\textrm{nl}}(n_{y}<0)]/2$.
\textbf{Right}: Bona fide ISHE and SGE nonlocal resistances obtained
by filtering of curves in the middle panel. Saturation fields: $B_{\textrm{sat},x}=0.2$\,\,T
and $B_{\textrm{sat},z}=1$\,\,T. Other parameters as in Fig. 2,
main text.}
\end{figure}

Figure \ref{fig:supp_1} shows the nonlocal resistance of the simulated
spin-valve device displaying ISHE and SGE. The bare signal for ``parallel''
($\hat{\boldsymbol{n}}=+\hat{y}$) and ``antiparallel'' ($\hat{\boldsymbol{n}}=-\hat{y}$)
initial configurations is shown in the left panel. The plateaus at
large field $|B|$ signal the saturation of the ferromagnetic contact.
The middle panel shows the spin-transresistance between parallel and
antiparallel configurations. The right panel shows the ISHE/SGE lineshapes
isolated using the protocol presented in the main text, i.e.,
\begin{equation}
\Delta R_{\textrm{SHE(SGE)}}=\frac{1}{2}\left.\left[\Delta R_{\textrm{nl}}(B)\pm\Delta R_{\textrm{nl}}(-B)\right]\right|\equiv\frac{1}{2}\left.\left[\Delta R_{\textrm{nl}}(\phi)\pm\Delta R_{\textrm{nl}}(-\phi)\right]\right|.\label{eq:protocol}
\end{equation}

The ISHE/SGE signal has even/odd parity with respect to mirror reflection
($\phi\rightarrow-\phi$). This shows that the filtering scheme accurately
separates the two independent components. Furthermore, the lineshapes
have the same exact form than those presented in Fig. 2, main text
(in which the tilting correction had been neglected by setting $R_{0}=$constant).

\newpage{}

\begin{figure}[H]
\centering{}\includegraphics[width=1\textwidth]{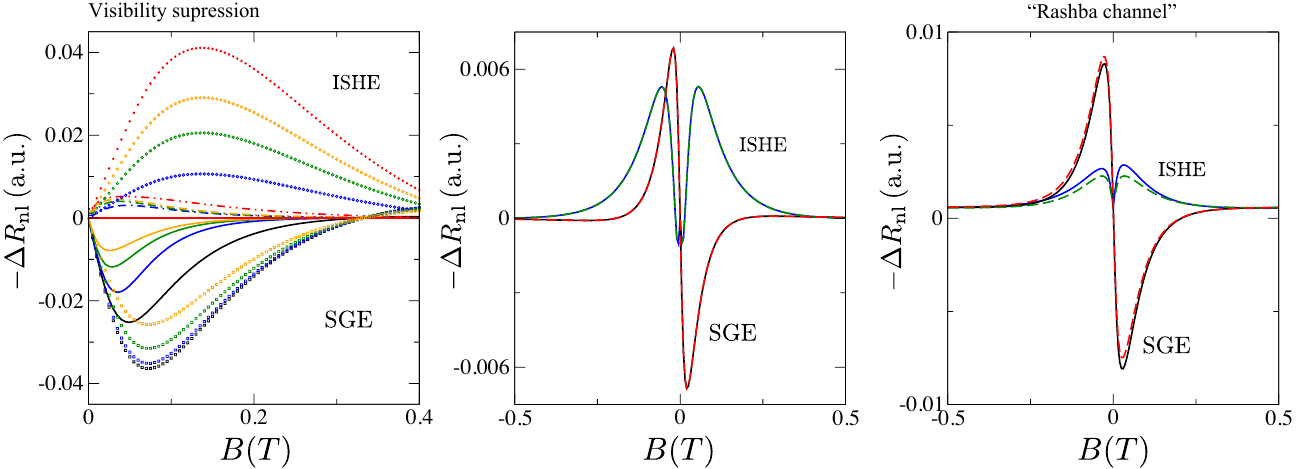} \caption{\label{fig:supp_2}\textbf{Left}: Calculated ISHE/SGE filtered nonlocal
resistance with magnetization tilting correction (solid lines) and
without (open symbols). The ratio of the two curves is precisely the
``visibility factor'' $\mathcal{V}(\mathbf{B})$ introduced in the
text. \textbf{Middle}: Simulation of individual ISHE and SGE nonlocal
signals (solid lines) and filtered signals using the protocol Eq.
(\ref{eq:protocol}) for a highly anisotropic spin channel with $l_{\perp}=10l_{\parallel}$.
\textbf{Right}: Same than middle panel but for a channel with $l_{\perp}=\frac{1}{2}l_{\parallel}$
and large Rashba precession parameter $l_{R}=5l_{s}$. Other parameters
as in Fig. \ref{fig:supp_1}.}
\end{figure}

Importantly, the unambiguous ISHE/SGE detection is possible even for
channels with anisotropic spin relaxation ($\tau_{\parallel}\neq\tau_{\perp}$),
i.e. devices where the spin channel itself is characterized by strong
SOC. This is shown in Fig.\ref{fig:supp_2} (left and middle panels).
The filtering is still accurate (albeit not \emph{exact}) in channels
with strong BR effect ($l_{R}$ a few times $l_{s}$) and breaks down
for $l_{R}\apprle l_{s}$. An example with $l_{R}=5l_{s}$ is shown
in Fig. Fig.\ref{fig:supp_2} (right panel). In graphene, this would
correspond to a proximity-induced BR field on the order of 1 meV.

\section*{S2. Hanle Precession: Role of BR field}

\begin{figure}[H]
\centering{}\includegraphics[width=0.35\columnwidth]{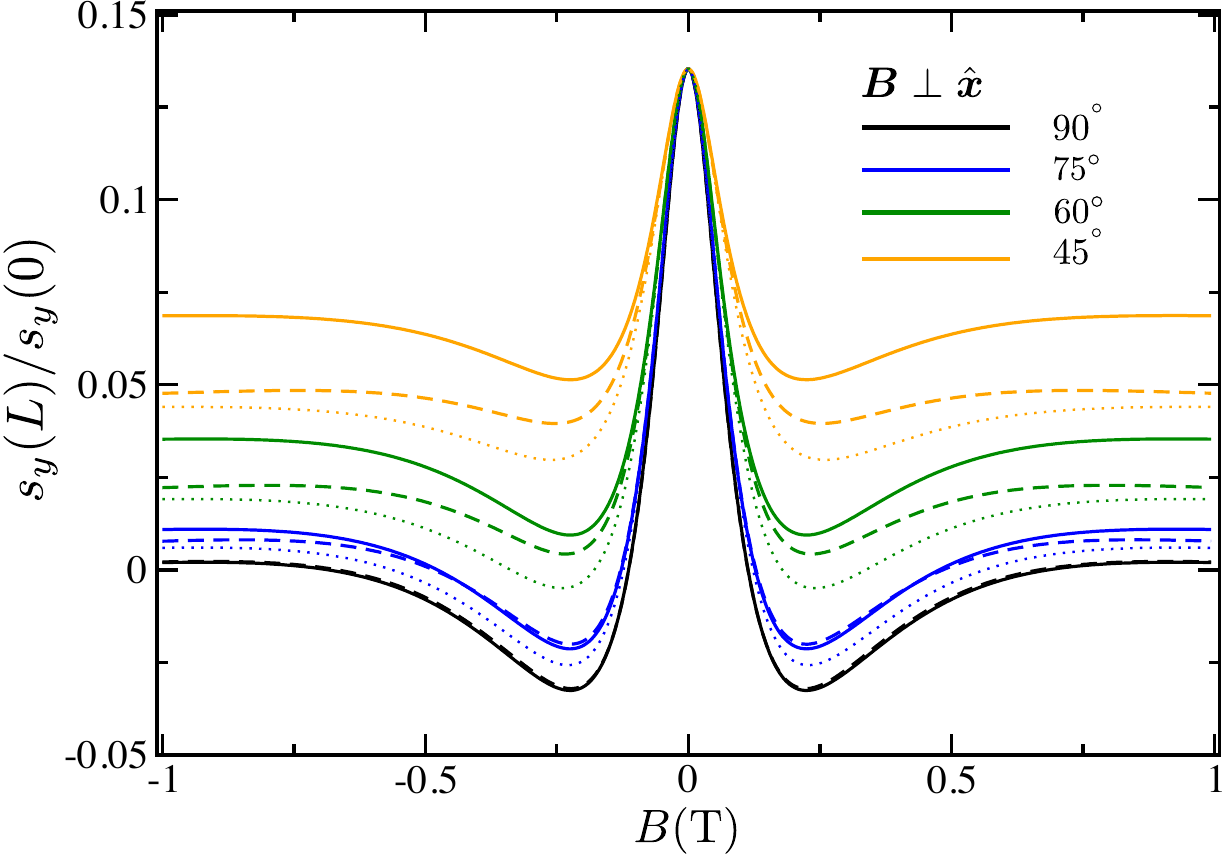}\caption{\label{fig:supp_03}Hanle precession induced by oblique magnetic field
$\boldsymbol{B}=(B\sin\phi\hat{\boldsymbol{z}}+|B|\cos\phi\hat{\boldsymbol{y}})$
for an isotropic channel (solid lines) and a Bychkov-Rashba channel
with $l_{R}=5l_{s}$ (dashed lines). The Hanle curve generated with
BR spin-orbit precession removed ($l_{R}=\infty$) is shown for comparison
(dotted line). The \textcolor{black}{injected spin polarization at
the contact is parallel to the ferromagnet easy-axis ($\mathbf{s}(x=0)\parallel\hat{y}$).}
Other parameters: $L=2l_{s}$ and $\tau_{s,\parallel}=0.1$\,ns.
Small tilting of electrode's magnetization neglected.}
\end{figure}

Here, we show that the coherent precession around a (strong) BR field
is also detectable in standard nonlocal Hanle measurements with spin-injector
(FM1) and spin-detector (FM2). Figure~\ref{fig:supp_03} \textcolor{black}{contrasts
the Hanle curve of an isotropic channel ($l_{R}=\infty$ and $l_{s,\parallel}=l_{s,\perp}=l_{s}$
\citep{Raes_SRTA_Graphene_16}) with that of a pure BR channel ($l_{R}<\infty$
and $l_{s,\parallel}=\sqrt{2}l_{s,\perp}$ \citep{GTMD_SRTA_Offidani_18}).
The impact of a short} $l_{R}\propto l_{s}$\textcolor{black}{{} is
perceptible for applied field with $\phi<90^{\circ}$, enabling diffusive
spins to precess in the $Oxz$ plane }\citep{comment_SP,ZhangWu_11}\textcolor{black}{.
}(This effect was previously noted in Ref.~\citep{ZhangWu_11}, where
it was \textcolor{black}{shown that the BR field induces damped oscillations
in $s^{a}(x)$ when injected spins are oriented in the $Oxz$-plane.)
Interestingly, the BR precession boosts the effective spin diffusion
length \citep{Lin_LongPaper}, which explains the enhanced $\hat{y}$-spin
density accumulation} away from the detector as compared to a hypothetical
BR channel with the same $l_{s,\parallel}=\sqrt{2}l_{s,\perp}$ but
with $l_{R}=\infty$.

\begin{thebibliography}{10}
\bibitem{Review_Interfaces_15}A. Manchon, H. C. Koo, J. Nitta, S.
M. Frolov, and R. A. Duine. New perspectives for Rashba spin--orbit
coupling. Nature Materials \textbf{14}, 871 (2015).

\bibitem{Review_Interfaces_16}A. Soumyanarayanan, N. Reyren, A. Fert,
C. Panagopoulos, Emergent phenomena induced by spin--orbit coupling
at surfaces and interfaces. Nature \textbf{539}, 509 (2016).

\bibitem{Review_Interfaces_17}F. Hellman, \emph{et al}., Interface-induced
phenomena in magnetism. Rev. Mod. Phys. \textbf{89}, 025006 (2017).

\bibitem{1stP_SCC_2DEGOxide}E. Lesne, \emph{et al}. Highly efficient
and tunable spin-to-charge conversion through Rashba coupling at oxide
interfaces. Nature Mater. \textbf{15}, 1261 (2016).

\bibitem{1stP_SCC_Ge-metal}S. Oyarzún, \emph{et al}. Evidence for
spin-to-charge conversion by Rashba coupling in metallic states at
the Fe/Ge(111) interface. Nature Comm. \textbf{7}, 13857 (2016).

\bibitem{1stP_SCC_metal_metaloxide}J. Puebla, \emph{et al}. Photoinduced
Rashba Spin-to-Charge Conversion via an Interfacial Unoccupied State.
Phys. Rev. Lett. \textbf{122}, 256401 (2019).

\bibitem{1stP_SCC_metal_TI}J.-C. Rojas-Sánchez, \emph{et al}. Spin
to Charge Conversion at Room Temperature by Spin Pumping into a New
Type of Topological Insulator: \textgreek{a} -Sn Films. Phys. Rev.
Lett. \textbf{116}, 096602 (2016).

\bibitem{1stP_BulkBReffect_BiTeI}K. Ishizaka, \emph{et al}. Giant
Rashba-type spin splitting in bulk BiTeI. Nature Mater. \textbf{1}0,
521 (2011).

\bibitem{GSOC_Rashba_09}E. I. Rashba, Graphene with structure-induced
spin-orbit coupling: Spin-polarized states, spin zero modes, and quantum
Hall effect. Phys. Rev. B 79, 161409 (2009).

\bibitem{GSOC_TMDs_Zhu_11}Z. Y. Zhu, Y. C. Cheng, and U. Schwingenschlögl.
Giant spin-orbit-induced spin splitting in two-dimensional transition-metal
dichalcogenide semiconductors. Phys. Rev. B \textbf{84}, 153402 (2011). 

\bibitem{GSOC_TMDs_Xiao_12}D. Xiao, G.-B. Liu, W. Feng, X. Xu, and
W. Yao, Coupled Spin and Valley Physics in Monolayers of MoS 2 and
Other Group-VI Dichalcogenides. Phys. Rev. Lett. \textbf{108}, 196802
(2012). 

\bibitem{GSOC_TMDs_Muniz_15}R. A. Muniz and J. E. Sipe, All-optical
injection of charge, spin, and valley currents in monolayer transition-metal
dichalcogenides. Phys. Rev. B\textbf{ 91}, 085404 (2015).

\bibitem{GSOC_Review_Garcia_18}J. H. Garcia, M. Vila, A. W. Cummings,
and S. Roche. Spin transport in graphene/transition metal dichalcogenide
heterostructures. Chem. Soc. Rev. \textbf{47}, 3359 (2018).

\bibitem{GTMD_Avsar_14}A. Avsar, J.Y. Tan, T. Taychatanapat, J. Balakrishnan,
G.K.W. Koon, Y. Yeo, J. Lahiri, A. Carvalho, A.S. Rodin, E.C.T. O\textquoteright Farrell,
G. Eda, A.H.C. Neto, and B. Ozyilmaz, Nat. Commun. \textbf{5}, 4875
(2014).

\bibitem{GTMD_Wang_15}Z. Wang, D.-K. Ki, H. Chen, H. Berger, A. H.
MacDonald, and A. F. Morpurgo, Strong interface-induced spin--orbit
interaction in graphene on WS2, Nat. Commun. \textbf{6}, 8339 (2015).

\bibitem{GTMD_Wang_16a}Z. Wang, \emph{et al.,} Origin and Magnitude
of \textquoteleft Designer\textquoteright{} Spin-Orbit Interaction
in Graphene on Semiconducting Transition Metal Dichalcogenides, Phys.
Rev. X \textbf{6}, 041020 (2016).

\bibitem{GTMD_Volkl_16b}T. Völkl, T. Rockinger, M. Drienovsky, K.
Watanabe, T. Taniguchi, D. Weiss, and J. Eroms, Magnetotransport in
heterostructures of transition metal dichalcogenides and graphene.
Phys. Rev. B \textbf{96}, 125405 (2017).

\bibitem{GTMD_Yang_17}B. Yang, M. Lohmann, D. Barroso, I. Liao, Z.
Lin, Y. Liu, L. Bartels, K. Watanabe, T. Taniguchi, and J. Shi, Strong
electron-hole symmetric Rashba spin-orbit coupling in graphene/monolayer
transition metal dichalcogenide heterostructures. Phys. Rev. B \textbf{96},
041409 (2017).

\bibitem{GTMD_Wakamura_18a}T. Wakamura, F. Reale, P. Palczynski,
S. Guéron, C. Mattevi, and H. Bouchiat, Strong Anisotropic Spin-Orbit
Interaction Induced in Graphene by Monolayer WS2. Phys. Rev. Lett.
\textbf{120} 106802 (2018).

\bibitem{GTMD_Omar_18b}S. Omar and B. J. van Wees, Spin transport
in high-mobility graphene on WS2 substrate with electric-field tunable
proximity spin-orbit interaction, Phys. Rev. B \textbf{97} 045414
(2018). 

\bibitem{GTMD_Optical_Luo_NanoLett2017}Y. K. Luo, \emph{et al. }Opto-Valleytronic
Spin Injection in Monolayer MoS2/Few-Layer Graphene Hybrid Spin Valves,
Nano Letters 17, 3877 (2017).

\bibitem{GTMD_Optical_Avsar_ACSNano2017}A. Avsar, \emph{et al.} Optospintronics
in Graphene via Proximity Coupling, ACS Nano\textbf{ 11,} 11678 (2017).

\bibitem{GTMD_SRTA_Cummings}A. W. Cummings, J. H. Garcia, J. Fabian,
and S. Roche. Giant Spin Lifetime Anisotropy in Graphene Induced by
Proximity Effects. Phys. Rev. Lett. \textbf{119}, 206601 (2017).

\bibitem{GTMD_SRTA_Ghiasi_17}T. S. Ghiasi, J. I.-Aynés, A. A. Kaverzin,
and Bart J. van Wees, Large Proximity-Induced Spin Lifetime Anisotropy
in Transition-Metal Dichalcogenide/Graphene Heterostructures, Nano
Lett. \textbf{17}, 7528 (2017).

\bibitem{GTMD_SRTA_Benitez18}L. A. Benítez, \emph{et al}. Strongly
anisotropic spin relaxation in graphene-transition metal dichalcogenide
heterostructures at room temperature, Nature Physics 14, 303 (2018).

\bibitem{GTMD_SRTA_Offidani_18}M. Offidani and A. Ferreira, Microscopic
theory of spin relaxation anisotropy in graphene with proximity-induced
spin--orbit coupling. Phys. Rev. B \textbf{98}, 245408 (2018). 

\bibitem{GSOC_Offidani_17}M. Offidani, M. Milletarí, R. Raimondi,
and A. Ferreira, Optimal Charge-to-Spin Conversion in Graphene on
Transition-Metal Dichalcogenides, Phys. Rev. Lett. \textbf{119}, 196801
(2017).

\bibitem{Milletari_ConservationLaws_17} M. Milletarí, M. Offidani,
A. Ferreira, and R. Raimondi, Covariant Conservation Laws and the
Spin Hall Effect in Dirac-Rashba Systems, Phys. Rev. Lett. \textbf{119},
246801 (2017).

\bibitem{GSOC_Huang_16}C. Huang, Y. D. Chong, and M. A. Cazalilla,
Direct coupling between charge current and spin polarization by extrinsic
mechanisms in graphene, Phys. Rev. B, \textbf{94}, 085414 (2016). 

\bibitem{Huang_NonLocalResistance_17}C. Huang, Y. D. Chong, and M.
A. Cazalilla, Anomalous Nonlocal Resistance and Spin-Charge Conversion
Mechanisms in Two-Dimensional Metals, Phys. Rev. Lett. \textbf{119},
136804 (2017).

\bibitem{AdGraph_Balakrishnan_13}J. Balakrishnan, G. K. W. Koon,
M. Jaiswal, A. H. Castro Neto \& B. Özyilmaz. Colossal enhancement
of spin--orbit coupling in weakly hydrogenated graphene. Nature Physics
\textbf{9}, 284 (2013).

\bibitem{AdGraph_Kaverzin_15}A. A. Kaverzin and B. J. van Wees. Electron
transport nonlocality in monolayer graphene modified with hydrogen
silsesquioxane polymerization. Phys. Rev. B \textbf{91}, 165412 (2015).

\bibitem{AdGraph_Wang_15}Y. Wang, X. Cai, J. R.-Robey, and M. S.
Fuhrer. Neutral-current Hall effects in disordered graphene. Phys.
Rev. B 92, 161411R (2015).

\bibitem{AdGraph_Volkl_19}T. Völkl, \emph{et al}. Absence of a giant
spin Hall effect in plasma-hydrogenated graphene. Phys. Rev. B\textbf{
99}, 085401 (2019).

\bibitem{SHE_SGE_2D_1}C. K. Safeer, \emph{et al}. Room-Temperature
Spin Hall Effect in Graphene/MoS2 van der Waals Heterostructures.
Nano Lett. 19, 1074 (2019).

\bibitem{SHE_SGE_2D_2}T. S. Ghiasi, A. A. Kaverzin, P. J. Blah and
B. J. van Wees. Charge-to-Spin Conversion by the Rashba--Edelstein
Effect in Two-Dimensional van der Waals Heterostructures up to Room
Temperature. Nano Lett. 19, 5959 (2019).

\bibitem{SHE_SGE_2D_3}L. A. Benitez \emph{et al}. Nat. Materials
\textbf{19}, 170 (2020).

\bibitem{SpinChannel_Tombros_07}N. Tombros, C. Jozsa, M. Popinciuc,
H. T. Jonkman, and B. J. van Wees, Electronic spin transport and spin
precession in single graphene layers at room temperature, Nature \textbf{448},
571 (2007).

\bibitem{SpinChannel_Kamalakar_15}M. V. Kamalakar, C. Groenveld,
A. Dankert, and S. P. Dash, Long distance spin communication in chemical
vapour deposited graphene. Nature Comm. 6, 6766 (2015). 

\bibitem{SpinChannel_Gurram_18}M. Gurram, S. Omar and B. J. van Wees,
Electrical spin injection, transport, and detection in graphene-hexagonal
boron nitride van der Waals heterostructures: progress and perspectives,
2D Mater. \textbf{5}, 032004 (2018).

\bibitem{SID_Reyren}N. Reyren, \textit{et al.} Gate-Controlled Spin
Injection at LaAlO3/SrTiO3 Interfaces. Phys. Rev. Lett. \textbf{108},
186802 (2012).

\bibitem{SID_Isasa}Miren Isasa, \textit{et al.} Origin of inverse
Rashba-Edelstein effect detected at the Cu/Bi interface using lateral
spin valves. Phys Rev B \textbf{93}, 014420 (2016).

\bibitem{SID_Kamerbeek}A. M. Kamerbeek, P. Hogl, J. Fabian and T.
Banerjee. Electric field control of Spin lifetimes in Nb-SrTiO\textsubscript{3}by
Spin-Orbit Fields. Phys. Rev. Lett \textbf{115}, 136601 (2015).

\bibitem{SID_Lou}X. Lou, \textit{et al.} Electrical detection of
spin transport in lateral ferromagnet--semiconductor devices. Nature
Physics \textbf{3},197 (2007).

\bibitem{ISHE_Valenzuela}S. O. Valenzuela and M. Tinkham. Direct
electronic measurement of the spin Hall effect. Nature 442, 176 (2006).

\bibitem{Raes_SRTA_Graphene_16}B. Raes, \emph{et al}., Determination
of the spin-lifetime anisotropy in graphene using oblique spin precession.
Nature Comm. \textbf{7}, 11444 (2016).

\bibitem{GTMD_SRTA_Ringer}S. Ringer, \emph{et al}. Measuring anisotropic
spin relaxation in graphene. Phys. Rev. B \textbf{97}, 205439 (2018).

\bibitem{Raes_anisotropic}B. Raes, \emph{et al}. Spin Precession
in anisotropic media. Phys. Rev. B 95, 085403 (2017).

\bibitem{Cummings_16}A. W. Cummings and S. Roche. Effects of Dephasing
on Spin Lifetime in Ballistic Spin-Orbit Materials. Phys. Rev. Lett.
116, 086602 (2016).

\bibitem{Lin_LongPaper}Y.-H. Lin, C. Huang, M. Offidani, A. Ferreira,
and M. A. Cazalilla. Theory of spin injection in two-dimensional materials
with proximity-induced spin-orbit coupling. Phys. Rev. B \textbf{100},
245424 (2019).

\bibitem{Tokatly_08}I.V. Tokatly. Equilibrium Spin Currents: Non-Abelian
Gauge Invariance and Color Diamagnetism in Condensed Matter. Phys.
Rev. Lett. 101, 106601 (2008).

\bibitem{Shen_PRB2014}K. Shen, R. Raimondi, and G. Vignale, Theory
of coupled spin-charge transport due to spin-orbit interaction in
inhomogeneous two-dimensional electron liquids. Phys. Rev. B \textbf{90},
245302 (2014).

\bibitem{Burkov_04}A. A. Burkov, Alvaro S. Núñez and A. H. MacDonald,
Theory of spin-charge-coupled transport in a two-dimensional electron
gas with Rashba spin-orbit interactions. Phys. Rev. B \textbf{70},
155308 (2004).

\bibitem{Tokatly_Sherman_10}I.V. Tokatly and E. Y. Sherman, Gauge
theory approach for diffusive and precessional spin dynamics in a
two-dimensional electron gas, Annals of Physics \textbf{325}, 1104
(2010).

\bibitem{RMP_SHE}J. Sinova, S. O. Valenzuela, J. Wunderlich, C.\LyXThinSpace H.
Back, and T. Jungwirth. Spin Hall effects. Rev. Mod. Phys. \textbf{87},
1213 (2015).

\bibitem{Gorini_15}C. Gorini, U. Eckern, and R. Raimondi. Spin Hall
Effects Due to Phonon Skew Scattering. Phys. Rev. Lett. \textbf{115},
076602 (2015).

\bibitem{SM}See Supplemental Material appended below, which includes
Ref. \citep{ZhangWu_11}.

\bibitem{ZhangWu_11}P. Zhang and M. W. Wu, Electron spin diffusion
and transport in graphene. Phys. Rev. B\textbf{ 84}, 045304 (2011).

\bibitem{Yang15}L. Yang \emph{et al}. Long-lived nanosecond spin
relaxation and spin coherence of electrons in monolayer MoS\_2 and
WS\_2. Nature Physics 11, 830 (2015). 

\bibitem{Min_06}H. Min, \emph{et al}. Intrinsic and Rashba spin-orbit
interactions in graphene sheets. Phys. Rev. B \textbf{74}, 165310
(2006).

\bibitem{Konschuh_10}S. Konschuh, M. Gmitra, and J. Fabian. Tight-binding
theory of the spin-orbit coupling in graphene. Phys. Rev. B \textbf{82},
245412 (2010).

\bibitem{Popovic05}D. Popovi\'{c}, \emph{et al}. High-resolution
photoemission on Ag/Au(111): Spin-orbit splitting and electronic localization
of the surface state. Phys. Rev. B \textbf{72}, 045419 (2005).
\end{thebibliography}
\end{document}